\documentclass[a4paper,conference]{IEEEtran/IEEEtran}
\def\ps@headings{
\def\@oddhead{\mbox{}\scriptsize\rightmark \hfil \thepage}
\def\@evenhead{\scriptsize\thepage \hfil \leftmark\mbox{}}
\def\@oddfoot{}
\def\@evenfoot{}}
\makeatother

\usepackage{pgfplots}
\usepackage{color}
\usepackage{subfig}
\usepackage{graphicx}
\usepackage{cite}
\usepackage{amssymb,amsfonts}
\usepackage{url}
\usepackage[utf8]{inputenc}
\usepackage{tabularx}
\usepackage{hhline}
\usepackage{multirow}
\usepackage{wrapfig}

\usepackage[outline]{contour}

\usepackage{pifont}
\usepackage{tikz}
  \usetikzlibrary{arrows,shapes,calc,positioning,shadows,trees,patterns,decorations}
  \tikzset{
    ncbar angle/.initial=90,
    ncbar/.style={
        to path=(\tikztostart)
        -- ($(\tikztostart)!#1!\pgfkeysvalueof{/tikz/ncbar angle}:(\tikztotarget)$)
        -- ($(\tikztotarget)!($(\tikztostart)!#1!\pgfkeysvalueof{/tikz/ncbar angle}:(\tikztotarget)$)!\pgfkeysvalueof{/tikz/ncbar angle}:(\tikztostart)$)
        -- (\tikztotarget)
    },
    ncbar/.default=0.5cm,
}
\tikzset{round left paren/.style={ncbar=0.5cm,out=110,in=-110}}
\tikzset{round right paren/.style={ncbar=0.5cm,out=70,in=-70}}

\usepackage{hyperref}
\usepackage{url}

\usepackage{amsmath}
\usepackage{amsmath,amsthm}
\usepackage{lmodern}
\usepackage{pgfplots}
\usepackage{algorithm}
\usepackage[noend]{algpseudocode}
\usepackage[export]{adjustbox}

\begin{document}

\title{Erasure code-based low storage blockchain node}

\author{
\IEEEauthorblockN{Doriane Perard, J\'er\^ome Lacan, Yann Bachy and Jonathan Detchart}
\IEEEauthorblockA{ISAE-Supaero, Universit\'{e} de Toulouse, France\\
firstname.name@isae-supaero.fr}
}

\maketitle

\begin{abstract}

The concept of a decentralized ledger usually implies that each node of a blockchain network stores the entire blockchain. However, in the case of
popular blockchains, which each weigh several hundreds of GB, the large amount of data to be stored can incite new or low-capacity nodes to run \textit{lightweight} clients. Such nodes do not participate to the global storage effort and can result in a centralization of the blockchain by very few nodes, which is contrary to the basic concepts of a blockchain.

To avoid this problem, we propose new \textit{low storage} nodes that store a reduced amount of data generated from the blockchain by using erasure codes. The properties of this technique ensure that any block of the chain can be easily rebuild from a small number of such nodes. 
This system should encourage low storage nodes to contribute to the storage of the blockchain and to maintain decentralization despite of a globally increasing size of the blockchain. This system paves the way to new types of blockchains which would only be managed by low capacity nodes.  
\end{abstract}

\section{Introduction}
\label{sec:intro}

One of the most interesting properties of blockchains is their decentralized nature allowing to avoid central authorities. 
Classically, each node participating to the blockchain must maintain it by participating to the consensus when inserting a new block and by storing the entire blockchain. However, the success of blockchains such as Bitcoin or Ethereum has highlighted a scalability problem. Indeed, the increasing size of these blockchains implies important efforts by medium capacity nodes: 

\begin{enumerate}
    \item CPU: each transaction has to be processed by every node. 
    \item Storage: the size of many popular blockchains is considerably increasing, which implicates an important storage effort for classic nodes.
    \item Network load: Points 1) and 2) can lead to a reduced number of nodes storing the entire blockchain. This could increase the network load for the remaining full nodes.

\end{enumerate}

Several solutions were recently proposed to cope with the general scalability problem of blockchains. For example, the Ethereum team~\cite{etherSharding:2017}  proposes a mechanism allowing to increase the blockchain's throughput. This solution consequently reduces the amount of nodes verifying each transaction without impacting the global security level. Doing so allows to process more transactions in parallel. Other solutions use auxiliary channels to debottleneck the network and save storage space, like Plasma Cash or lightning network~\cite{lightningNetwork:2016, plasma:2017}.

Considering the storage scalability problem, the most frequent solution consists in nodes running a lightweight client, thereby reducing their required storage effort. A side effect of this solution is an important network and processing load on full nodes satisfying light client requests.

From the storage point of view, it makes sense to compare blockchains with distributed storage systems like Cloud or Peer-to-peer (P2P) systems. Indeed, such systems also store large amounts of data replicated on several 
servers. All these systems do not use the same solutions but some global strategies can be observed. For highly requested (\emph{hot-}) data, replication is the most used solution. But for low requested (\emph{cold-}) data, erasure coding ~\cite{erasureCodeSurvey:2013} is often used. The interest of this solution is the capacity of improving the trade-off between storage effort, availability and reliability.

To the best of our knowledge, blockchains only use data replication. 
The aim of this paper is to propose the use of erasure codes in order to store the blockchain. To do so, we introduce a new type of node, an erasure code-based low storage node, which stores coded fragments for each block of the blockchain. The main benefit is to allow a reduction of the storage effort required by each node. This should encourage nodes with low storage capacity to participate to the storage effort of the blockchain and thus result in two other benefits which are 1) the contribution to maintaining the decentralized property of the blockchain and 2) the reduction of network load on each node by multiplying the number of nodes.

This paper is structured as follows. First, Section~\ref{sec_rel_work} describes some related works concerning blockchain storage mechanisms and erasure codes for distributed storage systems. Then, Sections~\ref{sec:description} and~\ref{sec:coding} respectively present a high level and low-level description of erasure-code based low storage nodes, the main contribution of this paper. After, Sections~\ref{sec:interest} and~\ref{sec:analysisParameters} present the interest of our low storage blockchain node and an analysis of the available parameters of our system. Section~\ref{sec:application} introduces some possible applications of our system. Finally, Section~\ref{sec:conclusion} concludes this paper and proposes some perspectives.

\section{Related works}
\label{sec_rel_work}
\begin{figure}[htb]
	\centering
	\resizebox{0.48\textwidth}{!}{

\begin{tikzpicture}[empty/.style={draw,thick,rounded corners,inner sep=.1cm}, to/.style={<->,>=stealth',semithick,font=\footnotesize}, dot/.style={dotted,semithick,font=\footnotesize}]

\draw(0,-0.8) node[empty,minimum width=100, minimum height=20] (n1) {$b_1$};
\draw(0,-1.6) node[empty,minimum width=100, minimum height=20] (n1) {$b_2$};
\draw(0,-2.4) node[empty,minimum width=100, minimum height=20] (n1) {$b_3$};
\draw(0,-3.2) node[empty,minimum width=100, minimum height=20] (n1) {$b_4$};
\draw(0,-4.0) node[empty,minimum width=100, minimum height=20] (n1) {$b_5$};
\draw(0,-4.8) node[empty,minimum width=100, minimum height=20] (n1) {$b_6$};
\draw(0,0) node[minimum width=100, minimum height=20] (n1) {Node 1};

\draw(4,-0.8) node[empty,minimum width=100, minimum height=20] (n1) {$b_1$};
\draw(4,-1.6) node[empty,minimum width=100, minimum height=20] (n1) {$b_2$};
\draw(4,-2.4) node[empty,minimum width=100, minimum height=20] (n1) {$b_3$};
\draw(4,-3.2) node[empty,minimum width=100, minimum height=20] (n1) {$b_4$};
\draw(4,-4.0) node[empty,minimum width=100, minimum height=20] (n1) {$b_5$};
\draw(4,-4.8) node[empty,minimum width=100, minimum height=20] (n1) {$b_6$};
\draw(4,0) node[minimum width=100, minimum height=20] (n1) {Node 2};

\draw(8,-0.8) node[empty,minimum width=100, minimum height=20] (n1) {$b_1$};
\draw(8,-1.6) node[empty,minimum width=100, minimum height=20] (n1) {$b_2$};
\draw(8,-2.4) node[empty,minimum width=100, minimum height=20] (n1) {$b_3$};
\draw(8,-3.2) node[empty,minimum width=100, minimum height=20] (n1) {$b_4$};
\draw(8,-4.0) node[empty,minimum width=100, minimum height=20] (n1) {$b_5$};
\draw(8,-4.8) node[empty,minimum width=100, minimum height=20] (n1) {$b_6$};
\draw(8,0) node[minimum width=100, minimum height=20] (n1) {Node 3};

\draw(12,-0.8) node[empty,minimum width=100, minimum height=20] (n1) {$b_1$};
\draw(12,-1.6) node[empty,minimum width=100, minimum height=20] (n1) {$b_2$};
\draw(12,-2.4) node[empty,minimum width=100, minimum height=20] (n1) {$b_3$};
\draw(12,-3.2) node[empty,minimum width=100, minimum height=20] (n1) {$b_4$};
\draw(12,-4.0) node[empty,minimum width=100, minimum height=20] (n1) {$b_5$};
\draw(12,-4.8) node[empty,minimum width=100, minimum height=20] (n1) {$b_6$};
\draw(12,0) node[minimum width=100, minimum height=20] (n1) {Node 4};

\end{tikzpicture}}
	\caption{Traditional blockchain}
	\label{fig:traditionalBC}
\end{figure}
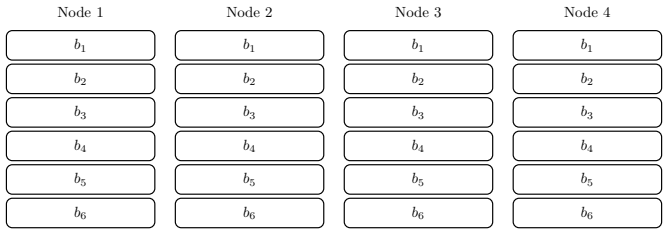

Let us consider blockchains as a sequence of variable sized blocks, virtually connected by their respective hashes and stored on a peer-to-peer network composed of several types of nodes. 
The first type, called \emph{full nodes}, stores the entire blockchain (cf. Fig.~\ref{fig:traditionalBC}). 
The second type, called \emph{lightweight nodes}, only stores the most recent blocks and the hashes of every block of the chain.  
This type of node allows to participate to a blockchain without storing the entire chain when the size of the blockchain becomes important.
In the rest of this section we will discuss three solutions allowing to face important data storage.
 
\subsection{Replicated systems}
\label{sec:replicatedSystem}
\begin{figure}[htb]
	\centering
	\resizebox{0.48\textwidth}{!}{

\begin{tikzpicture}[empty/.style={draw,thick,rounded corners,inner sep=.1cm}, to/.style={<->,>=stealth',semithick,font=\footnotesize}, dot/.style={dotted,semithick,font=\footnotesize}]

\draw(0,0.7) node[empty,minimum width=100, minimum height=20] (n1) {$b_1$};
\draw(0,1.5) node[empty,minimum width=100, minimum height=20] (n1) {$b_3$};
\draw(0,2.3) node[empty,minimum width=100, minimum height=20] (n1) {$b_5$};

\draw(0,3.0) node[minimum width=100, minimum height=20] (n1) {Node 1};

\draw(4,0.7) node[empty,minimum width=100, minimum height=20] (n1) {$b_6$};
\draw(4,1.5) node[empty,minimum width=100, minimum height=20] (n1) {$b_4$};
\draw(4,2.3) node[empty,minimum width=100, minimum height=20] (n1) {$b_2$};

\draw(4,3.0) node[minimum width=100, minimum height=20] (n1) {Node 2};

\draw(8,0.7) node[empty,minimum width=100, minimum height=20] (n1) {$b_1$};
\draw(8,1.5) node[empty,minimum width=100, minimum height=20] (n1) {$b_2$};
\draw(8,2.3) node[empty,minimum width=100, minimum height=20] (n1) {$b_3$};

\draw(8,3.0) node[minimum width=100, minimum height=20] (n1) {Node 3};

\draw(12,0.7) node[empty,minimum width=100, minimum height=20] (n1) {$b_4$};
\draw(12,1.5) node[empty,minimum width=100, minimum height=20] (n1) {$b_5$};
\draw(12,2.3) node[empty,minimum width=100, minimum height=20] (n1) {$b_6$};

\draw(12,3.0) node[minimum width=100, minimum height=20] (n1) {Node 4};

\end{tikzpicture}}
	\caption{Distributed replication blockchain}
	\label{fig:bcReplicated}
\end{figure}
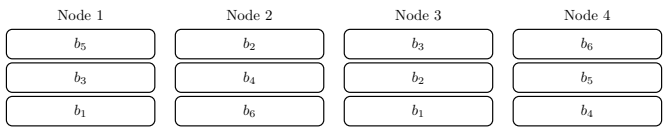

A first solution allowing to reduce the required storing capacity of every node consists in storing blocks among the nodes, in such a way that every block continues to be sustained by several nodes but not by all. A naive implementation of this solution is presented in Fig.~\ref{fig:bcReplicated}.

If one of the nodes storing a block becomes unavailable, the block continues to be stored by others. However, the global availability of this block is drastically reduced.
In this situation, malicious people could be tempted to target parts of the blockchain that are poorly replicated.

\subsection{Light nodes}
\label{sec:lightClient}
A second solution to storage limitation is offered by light clients. 
Many blockchains define their own type of light client~\cite{etherLightClient:2017,parityFast:2018,electrum:2017,bitcoinCorePruning:2016}.
Light clients do not store the entire blockchain, but only parts of it, generally the most recent blocks. As an example we can cite Simplified Payment Verification (SPV) which was originally designed by bitcoin's creator in his whitepaper~\cite{nakamoto2012bitcoin}. SPV allows nodes with limited resources to participate to a blockchain and perform payment verification, while only storing block headers. The node can not verify transactions by itself, but it requests block-content when necessary and uses the merkle tree to confirm the integrity of the data it received. 

Nodes running light clients present many problems:
\begin{itemize}
\item They don't store every block, so availability is not improved. If all nodes would run such clients, the overall blockchain availability would be endangered.
\item They never verify the entire blockchain, but instead rely their trust on other full nodes.
\end{itemize}

\subsection{Erasure code and distributed storage systems}
\label{sec:erasureCodeDSS}
A third solution to data storage is inspired from classic distributed storage systems.
Such systems often use erasure codes, like in  RAID (Redundant Array of Independent Disks)~\cite{Plank:1997:tutorialRAID}, Clouds~\cite{erasureCodeSurvey:2013} or P2P systems~\cite{erasureCodeAvailability}. The main idea is to encode the data into coded chunks which are distributed over several nodes of the system. 
The Storj system \cite{storj} uses a similar concept by storing coded and encrypted data in a P2P network that can be connected to a blockchain. The Swarm project~\cite{swarmOrangePaper:2016} also uses erasure codes in a blockchain context, it's purpose is to provide a decentralized and redundant store for dapp code and data, on the ethereum blockchain. 

In all these systems, erasure coding improves the trade-off between data availability and the amount of stored data. 

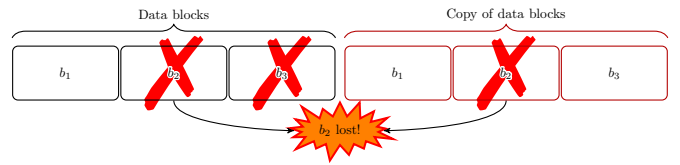
\begin{figure}[htb]
	\centering
	\contourlength{1pt}
\resizebox{0.48\textwidth}{!}{

\begin{tikzpicture}[empty/.style={draw,thick,rounded corners,inner sep=.0cm}, to/.style={->,>=stealth',semithick,font=\footnotesize}, dot/.style={dotted,semithick,font=\footnotesize}]

\draw(5,0) node [color=red] { \resizebox{!}{50pt}{\ding{55}}  };

\draw(7.8,0) node [color=red] { \resizebox{!}{50pt}{\ding{55}}  };

\draw(13.6,0) node [color=red] { \resizebox{!}{50pt}{\ding{55}}  };

\draw(2.2,0) node[empty,minimum width=78, minimum height=40] (n1) {\contour{white}{$b_1$}};
\draw(5.0,0) node[empty,minimum width=78, minimum height=40] (n2) {\contour{white}{$b_2$}};
\draw(7.8,0) node[empty,minimum width=78, minimum height=40] (n3) {\contour{white}{$b_3$}};
\draw(10.8,0) node[empty,minimum width=78, minimum height=40, draw=red!70!black] (n4) {\contour{white}{$b_1$}};
\draw(13.6,0) node[empty,minimum width=78, minimum height=40, draw=red!70!black] (n5) {\contour{white}{$b_2$}};
\draw(16.4,0) node[empty,minimum width=78, minimum height=40, draw=red!70!black] (n6) {\contour{white}{$b_3$}};

\draw[decorate,decoration={brace,raise=0.2cm,amplitude=10pt},thick]
(n1.north west) --node[midway,yshift=23pt,align=left,color=black!30!black,minimum width=80pt](db){Data blocks} (n3.north east) ;

\draw[decorate,decoration={brace,raise=0.2cm,amplitude=10pt},thick, color=red!70!black]
(n4.north west) --node[midway,yshift=23pt,align=left,color=black!50!black,minimum width=80pt]{Copy of data blocks} (n6.north east) ;

\draw(9.3,-1.5) node[starburst, draw, minimum width=3cm, minimum height=1cm,red,fill=orange,line width=1.5pt,text=black](boom)
{$b_2$ lost!};
\draw[to,thick] (n2.south) to[out = 270, in = 180, looseness = 0.5] (boom);
\draw[to,thick] (n5.south) to[out = 270, in = 0, looseness = 0.5] (boom);

\end{tikzpicture}}
	\caption{Traditional system}
	\label{fig:bcCoded1}
\end{figure}

To describe the concept behind erasure codes, let us first consider a replicated system, represented in Fig.~\ref{fig:bcCoded1}, which stores $3$ data blocks $b_1$, $b_2$ and $b_3$, these blocks are simply replicated for redundancy. 
Here, if a block and its replicated block are deleted or at least unavailable at a given time, this block can no longer be recovered. 

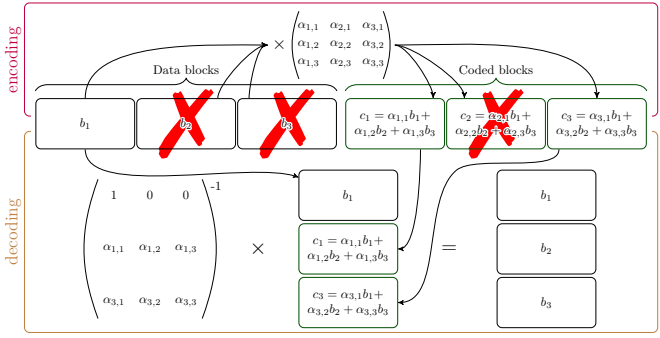
\begin{figure}[htb]
	\centering
	\contourlength{1pt}
\resizebox{0.48\textwidth}{!}{

\begin{tikzpicture}[empty/.style={draw,thick,rounded corners,inner sep=.0cm}, to/.style={->,>=stealth',semithick,font=\footnotesize}, dot/.style={dotted,semithick,font=\footnotesize}]

\draw(9.3,-4.8) node[anchor=south,empty, minimum width=500, minimum height=90,color=purple] (z1) {};
\draw(9.3,-4.8) node[minimum width=490, minimum height=5, fill=white] (z1) {};
\draw(9.3,-5.2) node[anchor=north,empty, minimum width=500, minimum height=160,color=brown] (z2) {};
\draw(9.3,-5.2) node[minimum width=490, minimum height=5, fill=white] (z21) {};
\draw(0.2,-5.2) node[anchor=east,minimum width=160, rotate=90,color=brown] {\Large decoding};
\draw(0.2,-4.8) node[anchor=west,minimum width=90, rotate=90,color=purple] {\Large encoding};

\draw(5,-5) node [color=red] { \resizebox{!}{50pt}{\ding{55}}  };

\draw(7.8,-5) node [color=red] { \resizebox{!}{50pt}{\ding{55}}  };

\draw(13.6,-5) node [color=red] { \resizebox{!}{50pt}{\ding{55}}  };

\draw(2.2,-5) node[empty,minimum width=78, minimum height=40] (c1) {$b_1$};
\draw(5.0,-5) node[empty,minimum width=78, minimum height=40] (c2) {\contour{white}{$b_2$}};
\draw(7.8,-5) node[empty,minimum width=78, minimum height=40] (c3) {\contour{white}{$b_3$}};
\draw(10.8,-5) node[empty,minimum width=78, minimum height=40, align=center, inner sep=.0cm, draw=green!30!black] (c4) {$c_{1}=\alpha_{1,1}b_{1}+$\\$\alpha_{1,2}b_{2}+\alpha_{1,3}b_{3}$};
\draw(13.6,-5) node[empty,minimum width=78, minimum height=40, align=center, inner sep=.0cm, draw=green!30!black] (c5) {\contour{white}{$c_{2}=\alpha_{2,1}b_{1}+$}\\\contour{white}{$\alpha_{2,2}b_{2}+\alpha_{2,3}b_{3}$}};
\draw(16.4,-5) node[empty,minimum width=78, minimum height=40, align=center, inner sep=.0cm, draw=green!30!black] (c6) {$c_{3}=\alpha_{3,1}b_{1}+$\\$\alpha_{3,2}b_{2}+\alpha_{3,3}b_{3}$};

\draw[decorate,decoration={brace,raise=0.2cm,amplitude=10pt},thick]
([yshift=0pt]c1.north west) --node[midway, yshift=23pt,align=left,color=black!30!black,minimum width=80pt]{Data blocks} ([yshift=0pt]c3.north east) ;

\draw[decorate,decoration={brace,raise=0.2cm,amplitude=10pt},thick, color=green!30!black]
(c4.north west) --node[midway,yshift=23pt,align=left,color=black!30!black,minimum width=80pt]{Coded blocks} (c6.north east) ;

\draw(8.4,-2.3) node[minimum width=25, minimum height=25] (m11) {$\alpha_{1,1}$};
\draw(8.4,-2.8) node[minimum width=25, minimum height=25] (m12) {$\alpha_{1,2}$};
\draw(8.4,-3.3) node[minimum width=25, minimum height=25] (m13) {$\alpha_{1,3}$};

\draw(9.3,-2.3) node[minimum width=25, minimum height=25] (m21) {$\alpha_{2,1}$};
\draw(9.3,-2.8) node[minimum width=25, minimum height=25] (m22) {$\alpha_{2,2}$};
\draw(9.3,-3.3) node[minimum width=25, minimum height=25] (m23) {$\alpha_{2,3}$};

\draw(10.2,-2.3) node[minimum width=25, minimum height=25] (m31) {$\alpha_{3,1}$};
\draw(10.2,-2.8) node[minimum width=25, minimum height=25] (m32) {$\alpha_{3,2}$};
\draw(10.2,-3.3) node[minimum width=25, minimum height=25] (m33) {$\alpha_{3,3}$};

\draw[thick]
 ([xshift=-5pt]m33.south east)to[round right paren] node[yshift=0pt](lme){} ([xshift=-5pt]m31.north east) ;
 \draw[thick]
  ([xshift=5pt]m13.south west)to[round left paren] node[xshift=-10pt](lm){\Large $\times$} ([xshift=5pt]m11.north west) ;

\draw[to,thick] (c1.north) to[out = 90, in = 180, looseness = 0.8] (lm);
\draw[to,thick] ([xshift=25pt]c2.north) to[out = 90, in = 180, looseness = 0.5] (lm);
\draw[to,thick] ([xshift=-30pt]c3.north) to[out = 90, in = 180, looseness = 0.5] (lm);

\draw[to,thick] (lme) to[out = 0, in = 90, looseness = 0.5] ([xshift=30pt]c4.north);
\draw[to,thick] (lme) to[out = 0, in = 90, looseness = 0.5] ([xshift=-25pt]c5.north);
\draw[to,thick] (lme) to[out = 0, in = 90, looseness = 0.8] (c6.north);

\draw(3,-7) node[minimum width=25, minimum height=25] (m11) {$1$};
\draw(3,-8.5) node[minimum width=25, minimum height=25] (m12) {$\alpha_{1,1}$};
\draw(3,-10) node[minimum width=25, minimum height=25] (m13) {$\alpha_{3,1}$};

\draw(4,-7) node[minimum width=25, minimum height=25] (m21) {$0$};
\draw(4,-8.5) node[minimum width=25, minimum height=25] (m22) {$\alpha_{1,2}$};
\draw(4,-10) node[minimum width=25, minimum height=25] (m23) {$\alpha_{3,2}$};

\draw(5,-7) node[minimum width=25, minimum height=25] (m31) {$0$};
\draw(5,-8.5) node[minimum width=25, minimum height=25] (m32) {$\alpha_{1,3}$};
\draw(5,-10) node[minimum width=25, minimum height=25] (m33) {$\alpha_{3,3}$};
 \draw[thick]

 (m33.south east)to[round right paren] node[yshift=50pt]{-1} (m31.north east) ;
 \draw[thick]
  (m13.south west)to[round left paren] (m11.north west) ;

\draw(9.5,-7) node[empty,minimum width=78, minimum height=40,color=black!30!black,align=center,inner sep=.1cm] (f1) {$b_1$};
\draw(9.5,-8.5) node[empty,minimum width=78, minimum height=40,draw=green!30!black,align=center,inner sep=.1cm] (f2) {$c_{1}=\alpha_{1,1}b_{1}+$\\$\alpha_{1,2}b_{2}+\alpha_{1,3}b_{3}$};
\draw(9.5,-10) node[empty,minimum width=78, minimum height=40,draw=green!30!black,align=center,inner sep=.1cm] (f3) {$c_{3}=\alpha_{3,1}b_{1}+$\\$\alpha_{3,2}b_{2}+\alpha_{3,3}b_{3}$};

\draw[to] (c1.south) to[out = 270, in = 160, looseness = 0.6] (f1);
\draw[to] ([xshift=20pt]c4.south) to[out = 270, in = 0, looseness = 0.5] (f2);
\draw[to] ([xshift=-30]c6.south) to[out = 270, in = 40, looseness = 0.6] (12.5,-6.3) to[out = 220, in = 0, looseness = 0.8] (f3);

\draw(15,-7.0) node[empty,minimum width=78, minimum height=40,align=center,inner sep=.1cm] (n1) {$b_{1}$};
\draw(15,-8.5) node[empty,minimum width=78, minimum height=40,align=center,inner sep=.1cm] (n1) {$b_{2}$};
\draw(15,-10) node[empty,minimum width=78, minimum height=40,align=center,inner sep=.1cm] (n1) {$b_{3}$};

\draw(12.3,-8.5) node[] {\LARGE =};
\draw(7,-8.5) node[] {\LARGE $\times$};

\end{tikzpicture}}
	\caption{Simple example of coded storage system}
	\label{fig:erasureExample}
\end{figure}

When using erasure codes, the $3$ redundancy blocks store linear combinations of the data blocks. 
With the same block unavailability, as shown in Fig.~\ref{fig:erasureExample}, we can observe that the $3$ blocks can be recovered from the $3$ available blocks by performing the inverse linear operations (we assume that the linear system is invertible).

Let's conclude that the use of erasure codes provides a better availability than classic data replication with the same amount of storage. The objective of this paper is to improve blockchain storage scalability by using erasure codes.

\section{Low storage nodes}
\label{sec:description}

In section \ref{sec_rel_work} we described two types of nodes, full nodes and lightweight nodes already deployed in many blockchains.

In this paper we introduce a new type of node, which is called \emph{Low Storage (LS) node}. They are described in the next paragraph. The way they recover a given block and the way they interact with the blockchain network is presented in the following paragraphs.

\subsection{Principle}
\label{sec:LSClient}
The main contribution of LS nodes to the blockchain system is to store only some \emph{coded fragments} of each block. These fragments are obtained by first splitting a block into fixed size fragments and then generating linear combinations of these fragments. Note that the combinations are specific to each block and to each node. 
This means that when several nodes store coded fragments from the same block, these will necessarily be different.

A detailed description of these operations is presented in Section~\ref{sec:coding}. The number of coded fragments stored by an LS node can vary according to different factors such as the age of the block or the storage capabilities of the node.

\subsection{Block recovery}
\label{sec:recovery}
When an LS node wants to recover a block from coded fragments stored by different LS nodes, it shall download coded fragments corresponding to the desired block upon several LS nodes. When the amount of coded fragments downloaded is more important than the amount of fragments the block was initially split into, the node can access the block by performing the inverse linear combinations.

\subsection{Integration of a new LS node in the network}
\label{sec:integration}
When an LS node wants to join the network, it needs to download all the blocks of the blockchain. This can be done from one or more trusted full nodes or from several LS nodes. For each block, it verifies its validity according to the previous block and generates its own coded fragments. Then, it removes the complete block and only stores the hash of the block and the coded fragments. Since the operations done on a block are independent from the other blocks, it can process the entire blockchain in sequence. This has the advantage of not needing to store the entire blockchain at once, therefore LS nodes are able to verify the entire blockchain with a reduced storage capacity.

\section{Coding the blockchain}
\label{sec:coding}

In this section we will present the technical aspects of coding block fragments. Firstly, we will discuss how we apply erasure coding to our system. The coding system in our system follows the same principle as coding mechanisms used in distributed storage systems (cf. Fig.~\ref{fig:erasureExample}). Secondly we will present how a node can recover a block based on coded fragments.

Before describing the coding operation, let us first define some notations. We denote by $N^{(i)}$ the nodes of the network, where $i$ is an unique identifier characterizing each node. We denote by $B^{(j)}$ the $j^{th}$ block of the blockchain. We consider that the first block is $B^{(0)}$. 

Let us denote by $s_B$ the maximum size of a block of the blockchain. Let us define two integers $k$ and $r$ respectively corresponding to the number of fragments of a block and the number of coded fragments stored by a node.
The choice of the values of $k$ and $r$ will be discussed in Section~\ref{sec:analysisParameters}.

\subsection{Coding the data}
\label{subsec:coding}

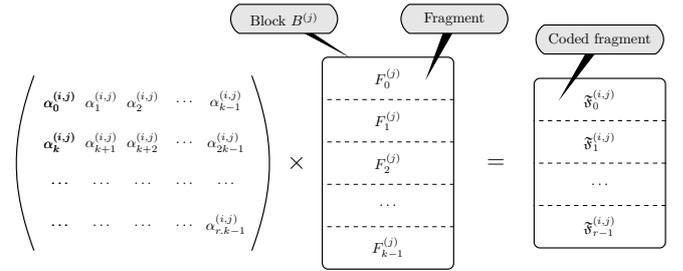
\begin{figure}[htb]
	\centering
	\resizebox{0.48\textwidth}{!}{
\begin{tikzpicture}[empty/.style={draw,thick,rounded corners,inner sep=.0cm}, to/.style={->,>=stealth',semithick,font=\footnotesize}, dot/.style={dotted,semithick,font=\footnotesize}]

\draw(7.9,-4.5) node[empty,minimum width=90, minimum height=80,rectangle split, rectangle split parts=5, rectangle split draw splits=false, rotate=0] (n1) {
\nodepart{one}
	\parbox[c][1cm]{40pt}{
		\begin{center}
			$F^{(j)}_{0}$
		\end{center}
	} 
\nodepart{two} 
	\parbox[c][1cm]{40pt}{
		\begin{center}
			$F^{(j)}_{1}$
		\end{center}
	} 
\nodepart{three} 
	\parbox[c][1cm]{40pt}{
		\begin{center}
			$F^{(j)}_{2}$ 
		\end{center}
	}
\nodepart{four}	
	\parbox[c][1cm]{40pt}{
		\begin{center} 
			$\ldots$
		\end{center}
	}
\nodepart{five}	
	\parbox[c][1cm]{40pt}{
		\begin{center} 

			$F^{(j)}_{k-1}$
		\end{center}
	}
};

 \draw[dashed] (n1.one split east) -- (n1.one split west);
  \draw[dashed] (n1.two split east) -- (n1.two split west);
   \draw[dashed] (n1.three split east) -- (n1.three split west);
      \draw[dashed] (n1.four split east) -- (n1.four split west);

\draw(0,-3) node[minimum width=35, minimum height=35] (m11) {$\alpha^{(i,j)}_{0}$};
\draw(0,-4) node[minimum width=35, minimum height=35] (m11b) {$\alpha^{(i,j)}_{k}$};
\draw(0,-5) node[minimum width=25, minimum height=35] (m12) {$\ldots$};
\draw(0,-6) node[minimum width=35, minimum height=35] (m13) {$\ldots$};

\draw(0,-3) node[minimum width=35, minimum height=35] (m11) {$\alpha^{(i,j)}_{0}$};
\draw(0,-4) node[minimum width=35, minimum height=35] (m11b) {$\alpha^{(i,j)}_{k}$};
\draw(0,-5) node[minimum width=25, minimum height=35] (m12) {$\ldots$};
\draw(0,-6) node[minimum width=35, minimum height=35] (m13) {$\ldots$};

\draw(1,-3) node[minimum width=25, minimum height=35] (m21) {$\alpha^{(i,j)}_{1}$};
\draw(1,-4) node[minimum width=25, minimum height=35] (m21b) {$\alpha^{(i,j)}_{k+1}$};
\draw(1,-5) node[minimum width=25, minimum height=35] (m22) {$\ldots$};
\draw(1,-6) node[minimum width=25, minimum height=35] (m23) {$\ldots$};

\draw(2,-3) node[minimum width=25, minimum height=35] (m31) {$\alpha^{(i,j)}_{2}$};
\draw(2,-4) node[minimum width=25, minimum height=35] (m31) {$\alpha^{(i,j)}_{k+2}$};
\draw(2,-5) node[minimum width=25, minimum height=35] (m32) {$\ldots$};
\draw(2,-6) node[minimum width=25, minimum height=35] (m33) {$\ldots$};

\draw(3,-3) node[minimum width=25, minimum height=35] (m31) {$\ldots$};
\draw(3,-4) node[minimum width=25, minimum height=35] (m31) {$\ldots$};
\draw(3,-5) node[minimum width=25, minimum height=35] (m32) {$\ldots$};
\draw(3,-6) node[minimum width=25, minimum height=35] (m33) {$\ldots$};

\draw(4,-3) node[minimum width=35, minimum height=35] (m41) {$\alpha^{(i,j)}_{k-1}$};
\draw(4,-4) node[minimum width=35, minimum height=35] (m41b) {$\alpha^{(i,j)}_{2k-1}$};
\draw(4,-5) node[minimum width=25, minimum height=35] (m42) {$\ldots$};
\draw(4,-6) node[minimum width=35, minimum height=35,align=center] (m43) {$\alpha^{(i,j)}_{r.k-1}$};
\draw[thick]
 (m43.south east)to[round right paren] node[yshift=50pt]{} (m41.north east) ;

 \draw[thick]
  (m13.south west)to[round left paren] (m11.north west) ;

\draw(10.5,-4.5) node[] {\LARGE =};

\draw(5.7,-4.5) node[] {\LARGE $\times$};

\draw(13,-4.5) node[empty,minimum width=90, minimum height=80,rectangle split,  rectangle split parts=4, rectangle split draw splits=false, rotate=0] (n1) {
\nodepart{one}
	\parbox[c][1cm]{40pt}{
		\begin{center}
			$\mathfrak{F}^{(i,j)}_{0}$
		\end{center}
	} 
\nodepart{two} 
	\parbox[c][1cm]{40pt}{
		\begin{center}
			$\mathfrak{F}^{(i,j)}_{1}$
		\end{center}
	} 
\nodepart{three} 
	\parbox[c][1cm]{40pt}{
		\begin{center}
			$\ldots$
		\end{center}
	} 
\nodepart{four} 
	\parbox[c][1cm]{40pt}{
		\begin{center}
			$\mathfrak{F}^{(i,j)}_{r-1}$
		\end{center}
	}
};
 \draw[dashed] (n1.one split east) -- (n1.one split west);
  \draw[dashed] (n1.two split east) -- (n1.two split west);
    \draw[dashed] (n1.three split east) -- (n1.three split west);

  \draw(13,-1.5) node[fill=black, rectangle callout, callout absolute pointer={(12,-3)}] {Coded fragment};
  \draw(13,-1.5) node[empty,fill=gray!20, rounded rectangle, text width=3cm, minimum height=20pt, align=center] {Coded fragment};

  \draw(9.5,-1.0) node[fill=black, rectangle callout, callout absolute pointer={(8.8,-2.5)}] {Fragment};
   \draw(9.5,-1.0) node[empty,fill=gray!20, rounded rectangle, text width=2.5cm, minimum height=20pt, align=center] {Fragment};

  \draw(5.4,-1.0) node[fill=black, rectangle callout, callout absolute pointer={(7.0,-1.95)}] {Block $B^{(j)}$};
    \draw(5.4,-1.0) node[empty,fill=gray!20, rounded rectangle, text width=2.5cm, minimum height=20pt, align=center] {Block $B^{(j)}$};

\end{tikzpicture}}
	\caption{Block coding process}
	\label{fig:bcCoded}
\end{figure}

When a node wants to code a block $B^{(j)}$, it must first split the block in $k$ fragments before applying a linear operation in order to obtain $r$ coded fragments (cf. Fig~\ref{fig:bcCoded}).

\begin{figure*}[h]
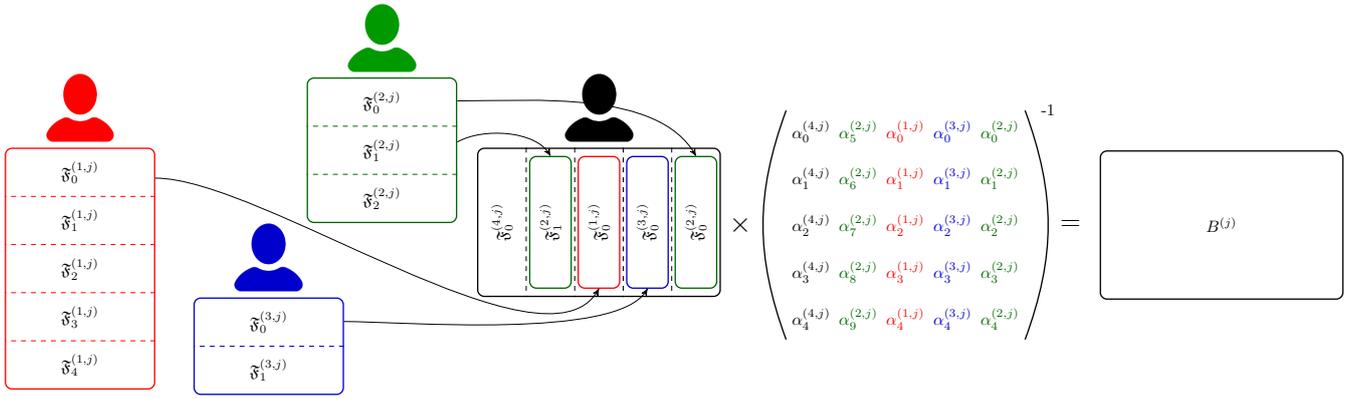

	\centering
	\include{fig_recover_block}
	\caption{Block recovery process}
	\label{fig:blRecover}
\end{figure*}

To generate $r$ coded fragments from block $B^{(j)}$, the node $N^{(i)}$ first splits the block 
into 
$k$ fragments $F^{(j)}_{0}, F^{(j)}_{1},\ldots,F^{(j)}_{k-1}$. 
If necessary, before splitting the block, it is padded with zeros up to $s_B$ in order to have a fixed fragment size $s_B/k$.  
Then, the node $N^{(i)}$ initializes a pseudo-random number generator with a seed defined as the concatenation of the binary expressions of $i$ and $j$ and generates the pseudo-random sequence $\{\alpha^{(i,j)}_{0}, \ldots, \alpha^{(i,j)}_{k.r-1} \}$ of $k.r$ coefficients in the finite field  $\mathbb{F}_{2^m}$. We propose to set $m:=8$ because the field $\mathbb{F}_{2^8}$ is generally considered as a good tradeoff between the complexity of its operations and the diversity it provides. This choice will be explained in Section~\ref{sec:analysisParameters}. In practical, the $256$ elements of this finite field are associated by a bijection to the $256$ bytes. 

To build the coded fragment $\mathfrak{F}^{(i,j)}_{u}$, where $0\leq~u~\leq~r-1$, the node considers the  $k$ coefficients $\{\alpha^{(i,j)}_{k.u}, \ldots, \alpha^{(i,j)}_{(k+1).u-1} \}$ and performs the following linear combination:
$$ \mathfrak{F}^{(i,j)}_{u} = \alpha^{(i,j)}_{k.u}.F^{(j)}_{0}  
+ \ldots, \alpha^{(i,j)}_{(k+1).u-1} .F^{(j)}_{k-1}$$

From a more practical point of view, if $f^{(j)}_{l,v}$ denotes the finite field element associated to the $v^{th}$ byte of the fragment $B^{(j)}_{l}$, then the $v^{th}$ byte of the coded fragment $\mathfrak{F}^{(i,j)}_{u}$ is defined by the finite field element $\mathfrak{f}^{(i,j)}_{u,v}$ computed as follows: 
$$ \mathfrak{f}^{(i,j)}_{u,v} = \alpha^{(i,j)}_{k.u}.f^{(j)}_{0,v}  
+ \ldots, + \alpha^{(i,j)}_{(k+1).u-1}.f^{(j)}_{k-1,v}$$ 
A more detailed presentation of the linear combinations of byte packets in finite fields is presented in~\cite{rizzo}.

Finally, the node removes the block $B^{(j)}$ and replaces it by the coded fragments $\mathfrak{F}^{(i,j)}_{0},\ldots,\mathfrak{F}^{(i,j)}_{r-1}$..

\subsection{Recovering the data}

Practically, a node $N^{(i)}$ downloads coded fragments $\mathfrak{F}^{(i_{0},j)}_{0},\ldots,\mathfrak{F}^{(i_{r-1},j)}_{r-1}$ from several  nodes $N^{(i_0)},\dots,N^{(i_{r-1})}$. Note that it is possible to request multiple coded fragments from the same node, as illustrated in Fig.~\ref{fig:blRecover}. Since the node $N^{(i)}$ knows the values of $i_0,\ldots,i_{r-1}$ and $j$, it is able to initialize the correct pseudo random number generator with the binary expressions of $i_0,\ldots,i_{r-1}$ and $j$ and thus can regenerate the coefficients $\alpha$ used when $N^{(i_0)},\dots,N^{(i_{r-1})}$ built their coded fragments. So, $N^{(i)}$ knows the $r$ linear equations used to build the $r$ coded fragments from the initial fragments. After downloading a sufficient number of coded fragments, the node will have enough equations to invert the linear system and recover the $k$ fragments (and thus the block) from the coded fragments. 

In Section~\ref{sec:analysisParameters} we will show that, with the field $\mathbb{F}_{2^8}$, in most cases the linear system can be inverted as soon as $k$ coded fragments are downloaded.

Note that non coded fragments can also be used in the decoding process. Indeed, they can be considered as a trivial linear combination of the fragments of the block (with $\alpha = 1$ for the corresponding fragment and $\alpha = 0$ for the other fragments).

\section{Low storage node interest}
\label{sec:interest}
The main objective of our system is to allow any node to contribute to an entire blockchain with a reduced storage effort. In this section we will discuss the different interests of our system compared to traditional blockchain nodes. Firstly we will compare the storage effort required by traditional vs coded blockchains. We will show how low storage blockchain nodes can relieve full nodes, and improve trust in the network. 
Secondly we will show how our system can improve the information security in terms of availability and integrity.

\subsection{Storage effort scalability}
\label{sec:sizes}

Let $c$ be the compression factor, defined by $c := k/r$. Note that a node that stores $r$ coded fragments must download $k-r$ extra coded fragments in order to recover a block. So, the higher the compression factor, the more coded fragments the node must download to be able to decode the entire block. The choice of these parameters will be discussed in section~\ref{sec:analysisParameters}.

When $r=k$, $c=1$ which means that the storage effort is not reduced. In this case, the data can be rebuild without downloading additional coded fragments. When $r$ decreases down to $1$, the compression factor increases up to $k$. Considering that each node must store at least one coded fragment per block, the maximum compression factor for each block is $k$.

One of the interests of our system is its scalability. Indeed, each node can adapt $r$ according to, for example, the age of a block by simply removing some of its stored coded fragments without re-calculating them.

Moreover, the number of coded fragments generated and stored on each node can independently be defined by each node, and should be adapted according to the desired storage effort of each node.

\subsection{Availability}

One of the main goals of our system is to improve the global availability and sustainability of a blockchain. In order to achieve this, our system reduces the storage effort required by a single node while participating to the global storage effort of the entire blockchain.

With our system we consider almost any node stores at least one coded fragment of every block. Thereby, even if each block is not stored in its initial form, every block is entirely stored among the different nodes, each one storing different fragments. In most cases, the initial block can be recovered by combining any $k$ coded fragments provided by any node of the network. 
For a system with a large amount of nodes, any node can leave the system or be unreachable without significantly impacting the availability.

To evaluate the availability, let us denote $n$ the number of nodes. Since each node can store a variable number of fragments, we note  $r_i$ the number of fragments stored by $i^{th}$ node. We assume that the $r_i$ can be modeled as realizations of an independent discrete random variable following a probability distribution $f$. Since a block can be recovered when at least $k$ coded fragments are available, the probability 
$\overline{P_c}$ of block irrecoverability from coded fragments is : 

 \makeatletter 
 \def\@eqnnum{{\normalsize \normalcolor (\theequation)}} 
  \makeatother

{ \small \begin{eqnarray} 
\overline{P_c} = p\left( \sum_{i=0}^{n-1} r_i < k \right) = \sum_{u=0}^{k-1} p \left( \sum_{i=0}^{n-1} r_i = u  \right) = \sum_{u=0}^{k-1} f^{*^n} (u)  
\label{eq:convolution_law} 
\end{eqnarray} }

where $f^{*^n} = \overbrace{ f*f*\cdots*f}^{n\ \textrm{times}}$ and $*$ the convolution between two distribution probabilities. 

To compare with a system based on random replication nodes (cf. Fig.~\ref{fig:bcReplicated}) with the same level of storage, we assume that each node stores each block with a probability of $1/c$. The probability that a block is \textit{not} stored by a node is thus $1-1/c$. By assuming that each node independently chooses which block to store, the probability $\overline{P_r}$ that a block is stored by no node at all is equal to:

\begin{equation}
\label{eq:random_distribution}
\overline{P_r} =  (1-\frac{1}{c})^n
\end{equation}

Let us compare these two systems with a numerical example. 
Let us assume that, in the coded system, the number $r$ of coded fragments stored by each node follows a geometric distribution $f_p(r) = p.(1-p)^{r-1}$. We then have:

\begin{equation}
\label{eq:convolution_geom_law}
	f_p^{*^n} (u) = \frac{(u-1)!}{(n-1)!(u-n)!}.p^n.(1-p)^{u-n} 
\end{equation}

\begin{figure}
	\centering
	\begin{tikzpicture}[y=7cm, x=0.28cm,font=\sffamily]
		\draw [->,xshift=0cm] (0,0) -- coordinate (x axis mid) (26,0);
		\draw [->,xshift=0cm] (0,0) -- coordinate (y axis mid) (0,0.22);
		\foreach \y in {0,0.025,...,0.2}{
				\draw [help lines, xshift=0cm] (0,\y) -- (25,\y);
			}
			
		\foreach \x in {0,5,...,25}
		\draw [xshift=0cm](\x,1pt) -- (\x,-3pt)
		node[anchor=north] {\x};
		\foreach \y in {0,0.1,0.2}
		\draw [xshift=0cm] (1pt,\y) -- (-3pt,\y)
		node[anchor=east] {\y};
		
		\node[below=0.5cm] at (x axis mid) {Number of stored fragments $r$};
		\node[rotate=90, shift={(0,0.14)}] at (y axis mid) {Probability};

		\draw[color=blue, xshift=0cm] plot[mark=-]
		file {data/geom_law_02.data};

		\begin{scope}[shift={(0,0.25)}]
			\draw[blue] (0,0) --
			plot[mark=-] (0.25,0) -- (0.5,0)
			node[right]{Geometric distribution};
		\end{scope}
	\end{tikzpicture}
	\caption{Probability for one node to store $r$ fragments, with geometric distribution of parameter $p = 0.2$}
	\label{fig:geometric_distribution}
\end{figure}
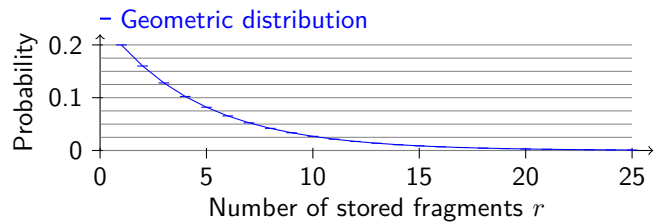

In our example, we assume that $k = 100$ and that the average compression factor of the block is $c = 20$. So, on average, each node stores $r = 5$ coded fragments. Since the mean of a geometric distribution of parameter $p$ is $1/p$, we have $p = 0.2$, as illustrated in Fig.~\ref{fig:geometric_distribution}. 

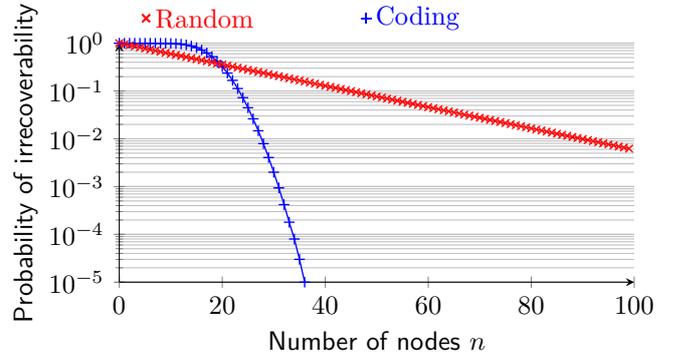
\begin{figure}
\begin{tikzpicture}
\centering
    \begin{semilogyaxis}[
        font=\sffamily,
        height=4.75cm,
        width=\linewidth-0.5cm,
        axis x line=bottom,
        axis y line=left,
        ymajorgrids,
        yminorgrids = true,
        ytick={0,0.00001,0.0001,0.001,0.01,0.1,1},
        xlabel = {Number of nodes $n$} ,
		ylabel = {Probability of irrecoverability},
		xmin = 0, xmax = 100, 
		ymin = 0.00001, ymax = 1 ]
        \addplot+[mark=+, line width=0.5pt] table[x index=0, y index=1] {data/proba_erasure_code_k_100.data};
        \addplot+[mark=x, line width=0.5pt] table[x index=0, y index=1] {data/proba_random_k_100.data};
        
    \end{semilogyaxis}

    \begin{scope}[shift={(0,2)}]
			\draw[yshift=1.5cm,xshift=0.1cm,red] (0,0) 
			plot[mark=x] (0.25,0)
			node[right]{Random};
			\draw[yshift=1.5cm,xshift=3cm, blue] (0,0) 
			plot[mark=+, mark options={fill=blue}] (0.25,0) 
			node[right]{Coding};
	\end{scope}
\end{tikzpicture}
	\caption{Probability one chosen block is irrecoverable, depending on $n$, with $k=100$ }
	\label{fig:proba_k_100}
\end{figure}

With these parameters, Fig.~\ref{fig:proba_k_100} shows the probability 
of block irrecoverability from coded fragments for the two systems, by using Eqs.~\ref{eq:convolution_law} and~\ref{eq:convolution_geom_law} for the coded system and Eq.~\ref{eq:random_distribution} for the random replication system.   

With the coded system and this probability law, starting from 37 nodes the probability to access a block equals 1 (with 5 digits precision), but with the randomly replicated it is only from 238 nodes. 

Regardless of the real probability law, with low storage nodes every node has to store at least $r=1$ fragments, so with only $k$ nodes we are sure to access the block.

Moreover, if we consider every node has the same probability of failure, using our mechanism, the overall probability of blockchain unavailability will drop. Indeed, considering the network storage capacity remains identical, our method distributes data over many more nodes, thereby distributing and reducing the global failure rate of the blockchain.

\subsection{Integrity}
This section will discuss how our system not only conserves the integrity of the blockchain but can improve trust by making data tampering more difficult.

\subsubsection{Hash-collision with coded fragments}
By construction, the integrity of a blockchain is guaranteed by the hash contained in each block. This hash validates the previous block of the chain and mechanically the entire blockchain. Therefore, any malicious node wanting to tamper a block of the blockchain would need to find a hash-collision in order to conserve the same hash result for the tampered block. 

Since the malicious node can only tamper some of the coded fragments used for the decoding process, he must figure out which correct fragments will be used for the decoding process. Then, he must search a hash-collision such that the result of the decoding process of the correct and the tampered fragments has the same hash as the initial block. This problem is at least as difficult as the classical hash-collision problem. So the use of erasure codes does not only reduce but rather increases the integrity of the system.  

\subsubsection{Malicious node identification} Erasure codes also provide solutions to identify malicious nodes tampering blocks. Tampered blocks will not be compatible and coherent with any set of $k$ valid coded fragments. This property allows each node to check the validity of other nodes and thereby improves the global security of the system. Moreover, this would allow the use of fraud proofs to report malicious nodes like detailed in~\cite{etherErasureCoding:2017}.

\subsubsection{Blockchain verification}
As explained in~\ref{subsec:coding}, the seed used by the pseudo-random number generator to generate the coefficients of the linear combination depends on the identity of the node. When decoding a block, each node uses the seeds belonging to the nodes from which the coded fragments were downloaded. When storing coded fragments of a block, these are necessarily re-coded with the local seed.

When a node wants to download coded fragments and decode the block, it uses $i$, the unique node identifier to regenerate the coefficients. 

The only way to obtain coded fragments coded with the correct coefficients, is to entirely code them from plain blocks. If a node does not follow this principle, the regenerated coefficients won't match and the decoding process will fail. This constraint forces each low storage node to verify the entire blockchain, which in turn will improve the overall integrity.

\subsection{Network load balancing}
We expect low storage nodes will contribute to a gain of the number of nodes, thereby improving the distribution of the blockchain over it's network.

As an example let us consider 1000 light nodes each connected with a 10Mbps network interface and only 5 full nodes serving the entire blockchain. In this situation each full node would have to be able to serve blocks at: $$\frac{10Mbps \times 1000}{5} = 2Gbps$$ in order to satisfy the 1000 light nodes.
With our method, if we consider each node stores 5 ($k=100$, $c=20$) coded fragments, with the same storage effort allowing 5 full nodes, our network can provide 100 ($5*c=5*20$) low storage nodes. In this situation each node would initiate 20 connections at 0.5Mbps (10Mbps / 20) which totals 20000 connections for all the nodes and therefore each full node would have to be able to serve blocks at only: $$\frac{\frac{10Mbps}{20}\times20000}{1000} = 100Mbps$$ in order to satisfy the 1000 nodes. This simple example clearly illustrates how our low storage nodes can decongest the network. Naturally connection establishment time and other overhead must also be considered when realizing a full network load study.

\section{Analysis of the parameters}
\label{sec:analysisParameters}

On each node, each block, and thus $k$ fragments, are replaced by $r$ coded fragments. One of the challenges is to determine $k$ and $r$, with the best compromise between compression and complexity. In this section, we will present some consequences when varying these parameters. Every blockchain and every type of node is different, it is therefore impossible to determine a generic value for $k$ and $r$ but a compromise must be found according to the system's constraints. Before discussing fragmentation and compression factor we will consider finite field size and processing complexity, since these two aspects will influence the choice of $k$ and $r$.

\subsection{Size of the finite field}
\label{sec:sizeFiniteField}
The choice of the finite field used to perform linear combinations mainly impacts two parameters: the probability of block recovery from downloaded coded fragments (which is better with a large finite field) and the encoding and decoding complexities (which is smaller with a small finite field). 

Reference~\cite{Studholme2010} shows that when the coefficients of the linear combinations are chosen in the field $\mathbb{F}_{2^m}$, the probability of recovering a block from $k+s$ downloaded fragments can be approximated by $1-2^{-m.(s+1)}$.  

Classically, for practical implementations, the possible sizes of the finite field $q$ are $2^{2^i}$, i.e. $2,\ 4,\ 16,\ 256,\ 65536$. Since extremely fast implementations can be developed for $2^m=256$ \cite{isa-l}, we propose to use this value in our system.

\subsection{Processing complexity}
\label{sec:procCimplexity}
The complexity of encoding consists in multiplying a $k \times r$-matrix by the $k$ original fragments. Then, there is ${k \times r \times s_B/k = r \times s_B }$ operations. So, the encoding complexity does not depend on $k$, but only on $r$. 

For decoding, the complexity consists in inverting a $k \times k$-matrix, and then multiplying it by the $k$ coded fragments. The pseudo-random matrix inversion has a complexity in $O(k^3)$. So the number of operations is ${O(k^3) + k^2 \times s_B/k} {= O(k^3) + k \times s_B} $ and thus depends only on $k$.
If the size of the block is large compared to $k$, then the matrix-vector multiplication ($k\times$ block size) is the most complex operation.

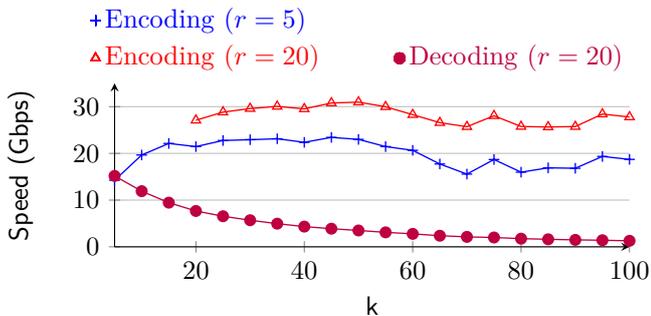
\begin{figure}
\begin{tikzpicture}
\centering
    \begin{axis}[
        font=\sffamily,
        height=3.75cm,
        width=\linewidth-0.5cm,
        axis x line=bottom,
        axis y line=left,
        ymajorgrids,
        xlabel = {k} ,
		ylabel = {Speed (Gbps)},
		xmin = 5, xmax = 100, 
		ymin = 0, ymax = 35 ]
		\addplot+[blue, mark=+, line width=0.5pt] table[x index=0, y index=1] {data/encoding_1M_r5.data};
        \addplot+[red, mark=triangle, line width=0.5pt] table[x index=0, y index=1] {data/encoding_1M_r20.data};
        \addplot+[purple, mark=*, mark options={fill=purple}, line width=0.5pt] table[x index=0, y index=1] {data/decoding_1M.data};
    \end{axis}
	    \begin{scope}[shift={(0,3)}]
            \draw[xshift=-0.5cm, blue] (0,0) 
			plot[mark=+] (0.25,0)
		    node[right]{Encoding ($r=5$)};

			\draw[xshift=-0.5cm, yshift=-0.5cm, red] (0,0)
			plot[mark=triangle] (0.25,0) 
		    node[right]{Encoding ($r=20$)};
		    
			\draw[yshift=-0.5cm,xshift=3.5cm, purple] (0,0)
			plot[mark=*, mark options={fill=purple}] (0.25,0) 
			node[right]{Decoding ($r=20$)};
	    \end{scope}

\end{tikzpicture}
\caption{Encoding and decoding speed, with block size~=~1Mb}	
\label{fig:encode_decode_1MB}
\end{figure}

Fig.~\ref{fig:encode_decode_1MB} shows the encoding and decoding speeds of 1MB blocks, with different values of $k$ and $r$. We used the erasure code implementation of \cite{isa-l} with avx2 instructions on a core i7-6700@3.2 GHz CPU. 
This graph allows to conclude that the processing cost is negligible. Indeed, coding speed is always greater than 15 Gbps for $r \leq 20$. 
Decoding speed is lower, but greater than 3.5 Gbps for $k \leq 50$ and equals to 1.5 Gbps for $k = 100$.

\subsection{Fragmentation}
As described in Section~\ref{subsec:coding}, each node, in order to generate its $r$ coded fragments, will split the initial block into $k$ fragments. The choice of this parameter can be different for each blockchain, but it must be the same for every user of the same blockchain. Table~\ref{tab:arg_k} shows the different arguments when choosing $k$. 

\renewcommand\tabularxcolumn[1]{m{#1}}
\newcolumntype{Y}{>{\centering\arraybackslash}X}
\begin{table}[htb]
\small
\caption{Choosing $k$}
\label{tab:arg_k}
\begin{center}
    
\resizebox{0.48\textwidth}{!}{
	\begin{tabularx}{0.5\textwidth}{|l||Y|}
		\hhline{-||-}
	  	 \textbf{Argument} & \textbf{$k$} \\
  		\hhline{=::=}
	 	Decoding complexity & $\searrow$ \\
	 	\hhline{-||-}
	 	 Compression factor flexibility~~ & $\nearrow$ \\
	 	 \hhline{-||-}
	 	 Max compression & $\nearrow$ \\
	 	 \hhline{-||-}
	\end{tabularx}
}
\end{center}
\end{table}

Overall we can conclude that $k$ should be chosen as big as possible, without exceeding 
$\frac{2^m}{2}=128$ ($2^m=256$, size of our finite field). 
If $s_B$, the maximum size of a block of the blockchain, is too small, it possible to group a small amount of blocks before coding them. Doing so will allow to increase $k$ and take a better benefit of low storage nodes.

\subsection{Compression factor}

The compression factor $c=k/r$ indicates by how much the storage effort of a node adopting the low storage mechanism can be reduced. Considering $k$ is system-wide determined, $c$ can only be chosen by varying $r$. Table~\ref{tab:arg_r} shows the different arguments when choosing $r$. 

\renewcommand\tabularxcolumn[1]{m{#1}}
\newcolumntype{Y}{>{\centering\arraybackslash}X}
\begin{table}[htb]
\small
\caption{Choosing $r$}
\label{tab:arg_r}
\begin{center}
    
\resizebox{0.48\textwidth}{!}{
	\begin{tabularx}{0.5\textwidth}{|l||Y|}
		\hhline{-||-}
	  	 \textbf{Argument} &\textbf{ $r$} \\
  		\hhline{=::=}
	 	 Blockchain availability & $\nearrow$ \\
	 	\hhline{-||-}
	 	 Regular block recovery~~ & $\nearrow$ \\
	 	\hhline{-||-}
	 	 Encoding complexity & $\searrow$ \\
	 	 \hhline{-||-}
	 	 High compression & $\searrow$ \\
	 	 \hhline{-||-}
	 	 Network load & $\nearrow$ \\
	 	 \hhline{-||-}
	\end{tabularx}
}
\end{center}
\end{table}

The choice of $r$ is up to the end user and will depend on the type of user. Choosing $r$ high will improve block recovery and reduce network load, also will this improve the overall blockchain availability as increasing $r$ increases the storage effort of a node. 
Choosing $r$ low, will reduce coding complexity and compression factor. Globally $r$ must be chosen according to the node's capacities. 

\section{Application}
\label{sec:application}
This section will discuss some possible applications of our system. Firstly we will study how our system can be adopted in existing blockchains. Secondly we will examine how our system could pave the way to blockchains on IoT devices.

\subsection{Positioning with existing system blockchains}
Low storage nodes can be the basis of new blockchains, but they can also be adapted in existing systems. Since our system only stores blocks in their coded form, any system adopting low storage nodes should not require a frequent access to old blocks in order to remain efficient.
In this part, we will explain how low storage nodes can be used in Ethereum and Bitcoin blockchains, and how to avoid abundant access to old blocks.  

\subsubsection{Ethereum}
\label{subsubsec:application:positioning:ethereum}
The Ethereum blockchain can be considered as a state transition system. Each state reflects the ownership status of all existing tokens at a given time. 
The Ethereum’s state machine begins with a genesis state. Each time new transactions are processed and validated, a new final state including these transactions is added to the system. At any time, this final state reflects the current state of the Ethereum blockchain. The current state is a single global truth, shared by every node. 

Each node stores the most recent state, therefore older blocks are not required to verify and process new transactions. 
In this situation, access to older blocks is only required by new nodes joining the network and wanting to verify the entire blockchain. Consequently, older blocks are rarely requested and can easily be coded without delaying the system.

\subsubsection{Bitcoin}
\label{subsubsec:application:positioning:bitcoin}
The Bitcoin protocol does not implement the exact same mechanism, but can also be seen as a state transition system. 
Considering new low-storage nodes need to verify the entire chain when joining the network, we could imagine to set up a similar state transition system as used by Ethereum.  

There have been several proposals to form a special tree to organize all unspent transaction outputs on the network, called an Unused Output Tree (UOT)~\cite{bitcoinCorePruning:2016,bitcoinLightClient:2016}. The transaction outputs are hashed as a Merkle tree, and the root hash is stored in the blockchain~\cite{bitcoinMTUT:2012}. This system allows to improve the security of SPV~\cite{nakamoto2012bitcoin}, because it is now possible to verify with the UOT that a transaction has not been spent in a subsequent transaction. 
In our case, it seems coherent to entirely store unspent transaction outputs, in a database for example, to have the equivalent of the current state of the system, as used by Ethereum.

\subsection{Internet of Things (IoT)}
Considering our system allows to drastically reduce storage effort required by a node without significantly increasing CPU sollicitation, our system could be an important step towards running blockchains on IoT devices. In order to examine this possibility we tested our system on a Raspberry Pi. Raspberry Pis are small singleboard computers using an ARM processor, they perfectly represent the processing capacity of an average IoT device or smartphone.

Table~\ref{tab:speed_raspberry} shows the encoding and decoding speeds of 1MB blocks, with different values for $k$ and $r$, using neon instructions on a Raspberry Pi 3b.
The algorithm used here does not implement parallel computing, so speeds could be easily improved by using the four threads of the quad-core ARM processor on our Raspberry Pi. 

Considering the encoding complexity is only related to $r$, we can observe the encoding speed is halved when $r$ is doubled. Therefore, we can estimate that the encoding speed will be roughly 100Mbps when $r = 40$. In the same way, decoding complexity is only related to $k$ and we can observe that the decoding speed is halved when $k$ is doubled.

The Raspberry Pi 3B we used integrates a 10/100 Mbps Ethernet port. We have shown that the encoding speed is above 100Mbps when $r \leq 40$, and with $k \leq 50$, the decoding speed is in the same order of magnitude. 
Therefore, when $k \leq 50$ and $r \leq 40$ the bottleneck of the system is the network, and not the encoding process.

Overall we can conclude that our system is perfectly compatible with low resource devices, while reducing the required storage effort.
 
\renewcommand\tabularxcolumn[1]{m{#1}}
\newcolumntype{Y}{>{\centering\arraybackslash}X}
\begin{table}[htb]
\small
\caption{Coding data speed on Raspberry Pi 3b, with block size~=~1MB, in Mbps}
\label{tab:speed_raspberry}
\begin{center}
    
\resizebox{0.48\textwidth}{!}{
	\begin{tabularx}{0.5\textwidth}{|c||Y|Y|Y|}
    \cline{2-4}
    \multicolumn{1}{c|}{} & K=25 & K=50 & K=100 \\ 
   \hhline{-=:=:=:}
    encoding, r=5 & 708 & 706 & 689 \\
    \hhline{|-||-|-|-|}
	encoding, r=10 & 362 & 363 & 355 \\ 
    \hhline{|-||-|-|-|}
    decoding & 146 & 82 & 43 \\ 
    \hhline{----}
	\end{tabularx}
}
\end{center}
\end{table}

\section{Conclusion}
\label{sec:conclusion}
The increasing popularity of blockchains reveals all the possibilities a decentralized ledger can offer. Nonetheless blockchains also have their limits and scalability rapidly appears problematic. 

In this paper, we introduce a new type of blockchain node, called Low Storage node, with the aim of proposing an alternative between light nodes and full nodes. This type of node uses erasure codes in order to allow users with reduced storage capacity to contribute to the blockchain, without storing the entire blockchain, but instead only linear combinated fragments of each block. Eventhough blocks are no longer stored in their original form, they can still be recovered by downloading fragments from other nodes and performing the inverse linear operation.

We have shown that our system can improve the overall availability of a blockchain without impacting data integrity. We believe our system will encourage new nodes to participate to the network, thereby relieving existing nodes. Moreover, CPU ressources required for the coding process are negligible when coding at the same speed as average home internet connections allow to download the blockchain.
We also tested our system on a nanocomputer, proving it's compatibility with low ressource IoT devices.

As future work we plan to further explore an IoT compatible blockchain model. We also plan to study how to exempt nodes from verifying the entire blockchain when joining the network.

\bibliographystyle{IEEEtran}
\bibliography{biblioBlockchains}

\end{document}